\documentstyle[prl,aps]{revtex}

\begin{document}

\draft

\preprint{\today}

\title{An Infrared study of the Josephson vortex state in high-T$_c$
cuprates}

\author{S.V.~Dordevic$^{1}$, Seiki Komiya$^{2}$,
Yoichi Ando $^{2}$, Y.J.~Wang$^{3}$, and D.N.~Basov$^{1}$}
\address{$^{1}$ Department of Physics, University of California,
San Diego, La Jolla, CA 92093}
\address{$^{2}$ Central Research Institute of Electric Power Industry,
Tokyo, Japan}
\address{$^{3}$ National High Magnetic Field Laboratory, Tallahassee, FL
32310}

\wideabs{

\maketitle

\begin{abstract}
We report the results of the c-axis infrared spectroscopy of
La$_{2-x}$Sr$_x$CuO$_4$ in high magnetic field oriented parallel
to the CuO$_2$ planes. A significant suppression of the superfluid
density with magnetic field $\rho_s(H)$ is observed for both
underdoped (x=0.125) and overdoped (x=0.17) samples. We show that
the existing theoretical models of the Josephson vortex state fail
to consistently describe the observed effects and discuss possible
reasons for the discrepancies.
\end{abstract}

}

\narrowtext

Although the mechanism of high-$T_c$ superconductivity in cuprates
still remains unresolved, significant progress has been made in
understanding of the mixed state of these systems
\cite{blatter94}. Many subtle predictions of the theories of the
vortex state have been experimentally verified
\cite{blatter94,yeh91,feinberg90,doyle93,parks95,gerrits95,dulic01,pimenov01,pesetski00,choi94,koshelev98,matsuda95,lake01,coffey91b,tachiki94,bulaevskii95,lilm97}.
In particular, electrodynamics in relatively small magnetic fields
(typically H$<$7 T) appears to be in accord with the models
proposed for both pancake vortices formed with ${\bf H}\perp$
CuO$_2$-planes, and for the Josephson vortices developing in the
${\bf H}\parallel$ CuO$_2$-plane configuration
\cite{yeh91,feinberg90,doyle93,parks95,gerrits95,dulic01}. In this
work we explored field-induced suppression of the superfluid
density in La$_{2-x}$Sr$_x$CuO$_4$ (LSCO) single crystals due to
Josephson vortices. By extending previous studies to stronger
fields (up to 17 T) we find marked departures of the experimental
data from conventional Josephson vortex theory. In particular the
superfluid density in underdoped LSCO appears to be much more
fragile than the models prescribe.  The implications of these
findings for our current understanding of cuprates will be
discussed.

We used infrared (IR) spectroscopy to examine the interlayer
(${\bf E}\parallel c$) response of high quality LSCO crystals
\cite{ando01}: an underdoped with x=0.125 (T$_c \simeq$ 32 K) and
a weakly overdoped with x=0.17 (T$_c \simeq$ 36 K). The
near-normal-incidence reflectance $R(\omega)$ was collected over a
broad temperature (6-300 K) and frequency (10-48,000 cm$^{-1}$)
range. In addition field-induced changes of the reflectivity
$R(H,6K)/R(0 T,6K)$ were measured under zero field cooling with
${\bf H} \parallel$ CuO$_2$, when vortices penetrate in-between
the CuO$_2$ planes \cite{tachiki94}. The uncertainty of the
absolute measurements in our apparatus does not exceed 0.5 $\%$,
whereas the relative errors of the field-induced changes are less
than 1 $\%$. The optical conductivity
$\sigma_{1}(\omega)+i\sigma_{2}(\omega)$ and the dielectric
function $\epsilon_{1}(\omega)+i\epsilon_{2}(\omega)$ were
calculated from $R(\omega)$ using Kramers-Kronig analysis.

In the superconducting state the real part of the optical
conductivity has two components:
$\sigma_1^{SC}(\omega)=\frac{\rho_s}{8}\delta(0)+\sigma_1^{reg}(\omega)$,
the first one due to superconducting condensate and the second due
to unpaired carriers below T$_c$ (a so called regular
contribution). The superfluid density $\rho_s$ is quantified with
the plasma frequency $\omega_s$:
$\rho_s=\omega_s^2=c^2/\lambda_c^2=4 \pi e^2 n_s/m^*$ related to
the density of superconducting carriers n$_s$ and their mass
m$^*$; $\lambda_c$ is the penetration depth. A delta function in
$\sigma_1^{SC}(\omega)$ gives rise to a term
$-(\omega_s/\omega)^2$ in $\epsilon_1(\omega)$. A common procedure
of extracting $\omega_s$ from $\epsilon_1(\omega)$ involves
fitting of the low energy part of the spectrum with 1/$\omega^2$.
The problem with this procedure is that it does not discriminate
screening due to superconducting condensate from regular
contribution. In order to fix this problem we use the following
correction procedure:

\begin{equation}
\epsilon_1(\omega)-\epsilon_1^{reg}(\omega)=-
\frac{\omega_s^2}{\omega^2}, \label{eq:lambda}
\end{equation}
where $\epsilon_1^{reg}(\omega)$ is the regular contribution to
dielectric function. In order to calculate
$\epsilon_1^{reg}(\omega)$ a KK-like transformation is employed:

\begin{equation}
\epsilon_1^{reg}(\omega)=1+\frac{2}{\pi}\int_{0^+}^{\infty}
\frac{\omega'\epsilon_2^{reg}(\omega')}{\omega'^2-\omega^2}d\omega',
\label{eq:sigma2}
\end{equation}
where $\epsilon_2^{reg}(\omega)$ is the regular contribution to
the imaginary part of dielectric constant, i.e. after a
$\delta(0)$-function has been subtracted. The procedure described
by Eq.~\ref{eq:lambda} and \ref{eq:sigma2} accounts for
contribution of unpaired carriers at $T<T_c$, but also phonons,
interband transitions, magnons, and all other finite energy
excitations.

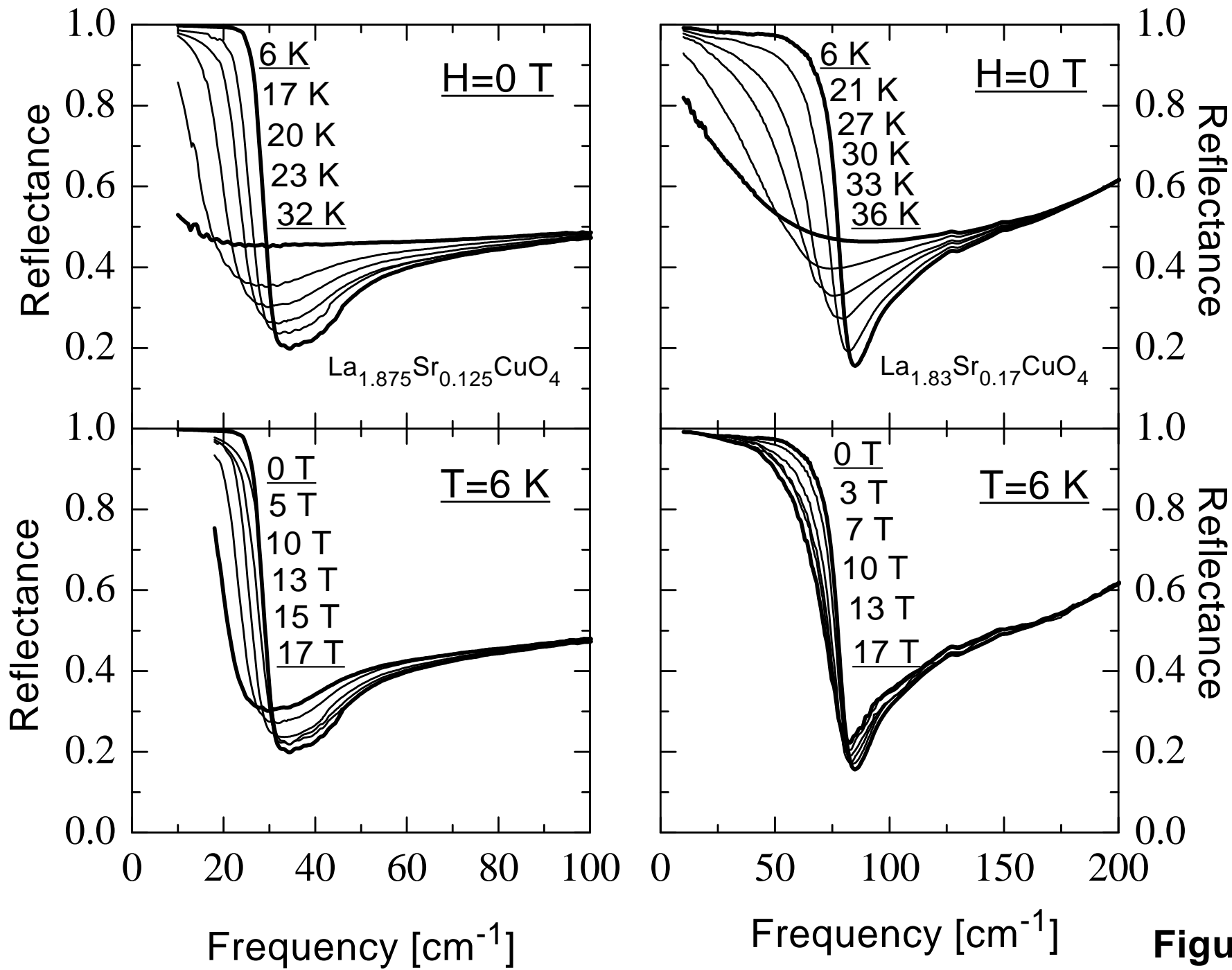
\begin{figure}
\caption{Infrared data for LSCO crystals with x=0.125 (left
panels) and x=0.17 (right panels). The top panels show c-axis
reflectance in zero field; the bottom panels: $R(\omega)$ in high
magnetic field.} \label{fig:ref}
\end{figure}

Figure \ref{fig:ref} displays the raw reflectance data. In zero
field the 6 K reflectance of both LSCO crystals is characterized
by a sharp plasma edge at the frequency
$\omega_s/\sqrt{\epsilon_{\infty}}$ ($\epsilon_{\infty}$ is the
high frequency dielectric constant). This form of reflectance,
which resembles a plasma edge in normal metals, is due to the zero
crossing in the real part of the dielectric function,
$\epsilon_1(\omega)=0$, and is commonly referred to as the
Josephson Plasma Resonance (JPR). As temperature increases, the
edge in $R(\omega)$  is smeared out and the minimum shifts to
lower frequency, indicating suppression of the superfluid density.
In the normal state the underdoped sample shows a very weak upturn
of $R(\omega)$ as $\omega\rightarrow 0$, indicating a small
metallic contribution at $T>T_c$. The upturn is stronger in the
x=0.17 crystal suggesting that the far IR conductivity increases
with doping. Similar enhancement  of the "metallic" trends in the
interlayer response is commonly found in other cuprate families.

Application of a magnetic field has a strong impact on the JPR
feature in both crystals (bottom panels): the plasma edges are
smeared and the minima in $R(\omega, H)$  also shift to lower
$\omega$. The plasma edge shift in 17 T field is as strong as 35
$\%$ in the x=0.125 sample and 10 $\%$ in the x=0.17 crystal. This
result is surprising given the fact that the strongest  field used
in our experiments is still orders of magnitude smaller than the
upper critical field H$_{c2}$ for the ${\bf H}\parallel$CuO$_2$
orientation.

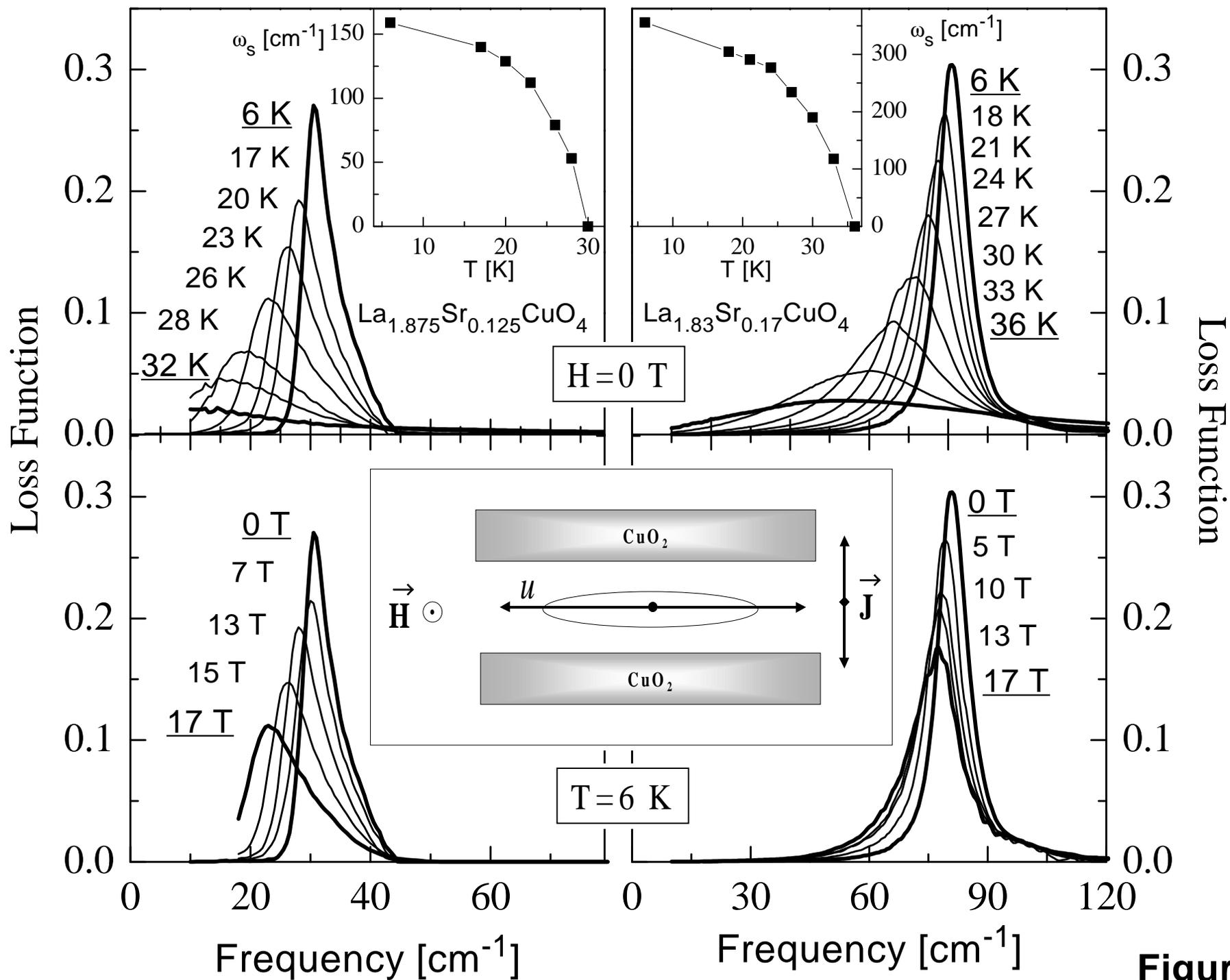
\begin{figure}
\caption{The loss function Im(1/$\epsilon(\omega)$) reveals
coupling to the longitudinal JPR mode. The top two panels show the
temperature dependence of the loss function and the bottom ones
the loss function in high magnetic field. The top insets show the
temperature dependence of the superfluid density and the bottom
one sketch the geometry of a Josephson vortex experiment.}
\label{fig:loss}
\end{figure}

It is instructive to present the evolution of the superconducting
condensate with H and T with the spectra of the loss function
$Im(1/\epsilon(\omega))=\epsilon_2(\omega)/
(\epsilon_1^2(\omega)+\epsilon_2^2(\omega))$
(Fig.~\ref{fig:loss}). At $T\ll T_c$  the loss function is peaked
at the frequency close to the JPR, whereas the width of the
$Im(1/\epsilon(\omega))$ mode  is proportional to the magnitude of
$\epsilon_2(\omega,T<T_c)$.  As temperature increases the peak
shifts to lower energies, indicating suppression of the superfluid
density. The insets in the top panels show the temperature
dependence of the superfluid density closely resembling the form
expected for a $d$-wave superconductor. At the lowest measured
temperature we found the following values of the superfluid
density: $\omega_s$=160 cm$^{-1}$ ($\lambda_c$ = 9.7 $\mu m$) for
12.5 $\%$ sample and $\omega_s$=360 cm$^{-1}$ ($\lambda_c$ = 4.3
$\mu m$) for 17 $\%$ material.

The behavior of the loss function in high magnetic field
(Fig.~\ref{fig:loss} bottom panels) is generally similar to the
H=0 data taken at finite temperature: the peak softens and its
width is enhanced as the field increases. Fig.3A quantifies the
demise of $\omega_s$ in magnetic field. At 17 T the superfluid
density is reduced by 38 $\%$ in the underdoped and 12 $\%$ in the
overdoped sample; these values are in full agreement with the
strength of the effect inferred directly from raw data in
Fig.~\ref{fig:ref}. We also note qualitatively different  form of
the $\omega_s(H)$ dependencies in the two samples. Two principal
mechanisms of the superfluid density suppression in high magnetic
field are: 1) direct pair-braking of the condensate and 2)
dissipation associated with vortex dynamics. The former process is
usually discarded in view of giant $H_{c2}$ values in the ${\bf
H}\parallel$ CuO$_2$ configuration. Within the latter picture the
oscillating electric field with the {\bf E}-vector along the
$c$-axis leads to a transverse motion $u$ of Josephson vortices
\cite{blatter94} located between the CuO$_2$ planes in the
direction perpendicular to the field (see the bottom inset of
Fig.~\ref{fig:loss}). Electrodynamics of Josephson vortices has
been thoroughly discussed in several papers.
\cite{coffey91b,tachiki94,bulaevskii95} Below we show that the
behavior of JPR in LSCO crystals cannot be fully understood within
the proposed models, especially in the underdoped sample.

Bulaevskii {\it et al} have worked out the following prediction
for the field dependence of $\omega_s$: \cite{bulaevskii95}
\begin{equation}
\omega_s(H)=\omega_0\left[1-\frac{\pi}{8}
\frac{H}{H_0}\ln\frac{H_0}{H}\right].
\label{eq:bulaevskii}
\end{equation}
The effect of magnetic field is directly related to the strength
of the coupling between the CuO$_2$ layers that is parameterized
through the characteristic field H$_0$=$\Phi_0/\gamma s^2$. In
this expression $\Phi_0$ is the flux quantum,
$\gamma=\lambda_c/\lambda_{ab}$ ($\lambda_{ab}$ is the in-plane
penetration depth) is the anisotropy factor and $s\simeq$ 1.32 nm
is the interlayer distance. We estimated H$_0$=36.7 T for the
x=0.125 with $\lambda_{ab}$ = 0.2 $\mu$m
(Ref.~\onlinecite{uchida96}) and H$_0$= 55 T for the x=0.17 sample
with $\lambda_{ab}$ = 0.3 $\mu$m (Ref.~\onlinecite{uchida96}). In
Fig.~3B we plot $\omega_s(H)/\omega_s(0)$ as a function of H/H$_0$
along with the theoretical dependence (full line). The overall
trends of $\omega_s(H)$ are clearly different for the two
crystals. In particular, the x=0.125 material reveals a dramatic
reduction of the condensate strength compared to the model
prediction for H/H$_0>$ 0.3. For x=0.17 sample such high magnetic
fields could not be achieved with the present experimental setup.

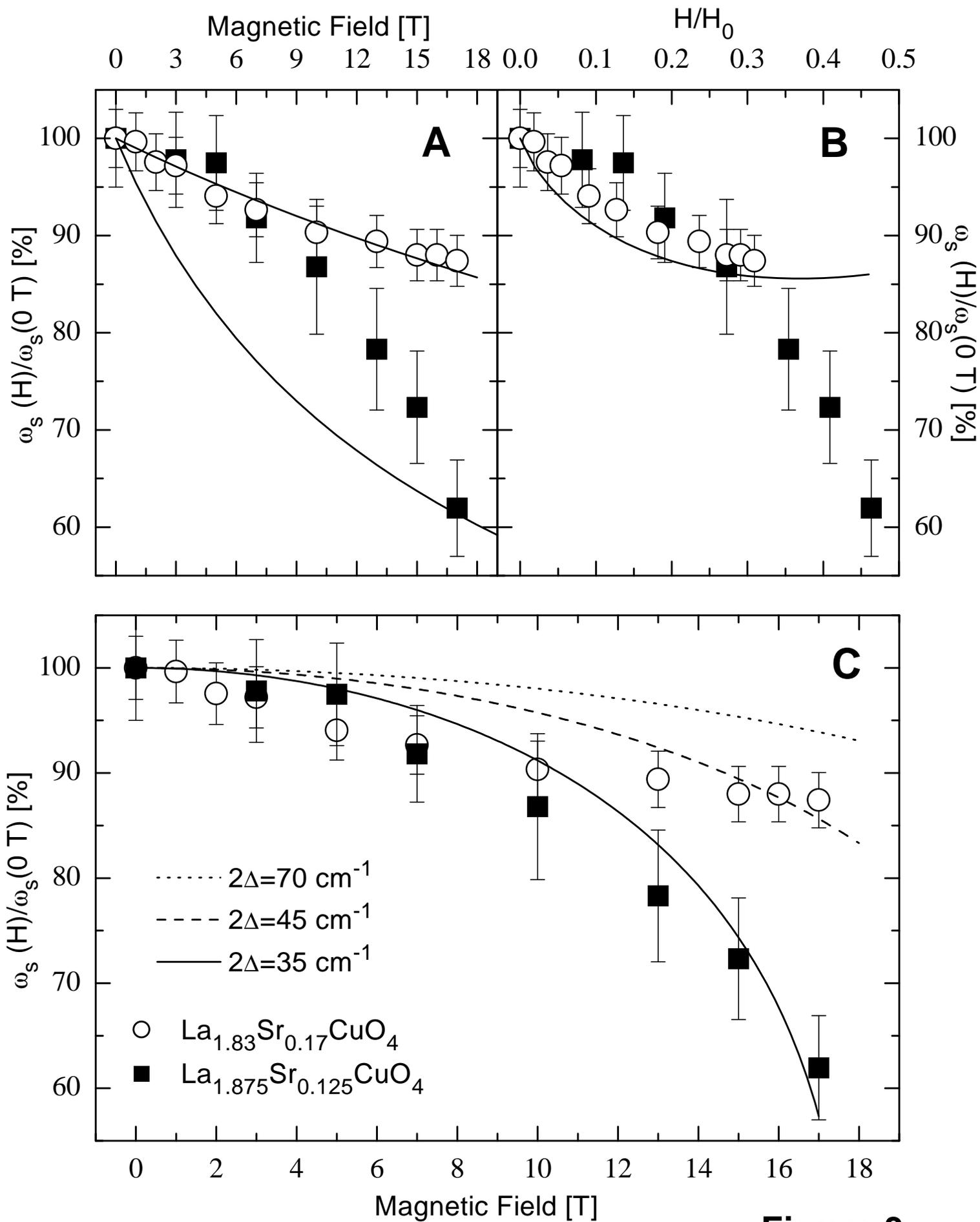
\begin{figure}
\caption{Change of the superfluid density with magnetic field
$\omega_s(H)/\omega_s(0)$ alone with the theoretical results
obtained for the TKT model (panel A), Bulaevskii {\it et al} model
(panel B) and scenario taking into account the nodal Zeeman effect
(panel C). While conventional models of the Josephson vortex state
are inconsistent with the high field data for the underdoped
x=0.125 crystal (A and B), plausible description can be achieved
within the picture discussed by Won {\it et al.}
\protect\cite{won} using the magnitude of the gap in from the
in-plane measurements for the same crystal \protect\cite{dumm02}.}
\label{fig:plasmon}
\end{figure}

We also attempted to describe $\omega_s(H)$ within a
phenomenological scenario of vortex dynamics
\cite{coffey91b,tachiki94}. Tachiki, Koyama and Takahashi (TKT)
\cite{tachiki94} obtained an explicit result for the complex
dielectric function of a layered superconductor in the mixed
state:
 \begin{equation}
\epsilon(\omega)=\epsilon_{\infty}-\frac{\frac{\omega_n^2}
{\omega^2+i\gamma_{sr}\omega}+\frac{\omega_s^2}{\omega^2}}{1+\frac{\phi_0}
{4\pi\lambda_c^2} \frac{H}{\kappa_p-i\eta\omega-M\omega^2}},
\label{eq:tkt}
\end{equation}
where M is the vortex inertial mass, $\eta$ is the viscous force
coefficient and $\kappa_p$ is the vortex pinning constant. In
Eq.~\ref{eq:tkt}, $\epsilon_{\infty}$ is the real part of the
dielectric function above the plasmon and $\omega_n$ and
$\gamma_{sr}$ are the regular component (due to unpaired carriers
below T$_c$) plasma frequency and scattering rate.
Eq.~\ref{eq:tkt} can be regarded as a H$\neq$0 generalization of
the well known "two fluid" model of superconductivity, commonly
used in the microwave and IR frequency ranges \cite{hosseini99}.
In order to model the data in Fig.~\ref{fig:plasmon} we extracted
the $\omega_s(H)$ behavior from Eq.~\ref{eq:tkt} by exploring
Re[$\epsilon(\omega)$] in the limit of $\omega\rightarrow 0$:
$\omega_s^2 = \omega^2\times \epsilon_1(\omega)$. Some of the
parameters of the TKT equation needed to obtain this result were
readily available from the fits of $R(\omega,6K,H=0)$:
$\epsilon_{\infty}$=27, $\omega_n$=200 cm$^{-1}$ (for 12.5 $\%$
sample) and 1,100 cm$^{-1}$ (for 17 $\%$ sample),
$\gamma_{sr}$=5,000 cm$^{-1}$. Similar to all previous
spectroscopic works \cite{parks95,gerrits95,wu90} we set the
vortex mass to zero  in Eq.~\ref{eq:tkt}. The viscous drag
constant $\eta$ can be calculated within Coffey-Clem approach
\cite{coffey91b} yielding $\eta$=7 Pa cm and 28 Pa cm, for the
underdoped and overdoped samples respectively. \cite{eta_values}
Therefore we are left with the pinning force constant $\kappa_p$
as the only fitting parameter in Eq.~\ref{eq:tkt}. As Fig.~3A
shows the theory gives a very good fit for the overdoped sample,
with $\kappa_p$= 6,000-11,000 Pa, a value comparable with that of
nearly optimally doped YBa$_2$Cu$_3$O$_{y}$ \cite{wu90} (in the
same field configuration). For the underdoped sample on the other
hand, we could not obtain a good fit for any value of $\kappa_p$.
In order to reproduce the overall depression of the superfluid
density in 17 T field we have to adopt $\kappa_p$=150-200 Pa. We
believe that such a vast difference in the magnitude of $\kappa_p$
between the two samples is another signal of inability of the TKT
scenario to account for the experimental situation at least in
underdoped LSCO crystals.

One obvious distinction between the underdoped and overdoped
samples is the different nature of the interlayer transport.
Earlier experiments \cite{boebinger96} established that the x=0.16
doping in the LSCO system separates the region of "insulating"
(for x$<$0.16) and "metallic" (x$>$0.16) ground states. This
result may have important implications for the nature of the ${\bf
H}\parallel$ CuO$_2$-vortices. The x=0.125 crystal is likely to
fall in the regime where a description within the formalism of
Josephson vortices is applicable  owing to the insulating
character of the "medium" separating the CuO$_2$ planes. However,
as doping progresses  to the over-doped side eventually the ${\bf
H}\parallel$ CuO$_2$-vortices will evolve into Abrikosov-type
vortices having a normal core.  Therefore our x=0.17 crystal may
be closer to the regime where additional factors involving complex
character of the vortex cores in $d$-wave superconductors have to
be taken into consideration. \cite{lake01,arovas97,han00}
Surprisingly, the Josephson analysis (Eqs.~\ref{eq:bulaevskii} -
\ref{eq:tkt}) obviously ignoring the latter issues, is less
problematic for the x=0.17 crystal but fails for the x=0.125
sample which, based on its "insulating" resistivity at
$T\rightarrow 0$, is likelier to comply with
Eqs.~\ref{eq:bulaevskii} - \ref{eq:tkt}. In particular, the TKT
analysis yields an unexpectedly small pinning constant which is
hard to reconcile with prominence of the intrinsic inhomogeneities
in underdoped cuprates, since inhomogeneities would normally
promote vortex pininnig.

When searching for reasons for the inability of
Eqs.~\ref{eq:bulaevskii} - \ref{eq:tkt} to consistently describe
the anomalous sensitivity of the superfluid response in underdoped
LSCO to magnetic field it is prudent to revisit the assumptions of
these models. Indeed, Eqs.~\ref{eq:bulaevskii} - \ref{eq:tkt} are
valid for Josephson coupling between $uniform$ $s$-wave
superconductors; validity of both assumptions for high-T$_c$
materials is questionable. One implication of the $d$-wave order
parameter (firmly established for cuprates) is that the Zeeman
energy associated with the ${\bf H}\parallel$ CuO$_2$ field can no
longer be disregarded for states close to the node. \cite{sondhi}
The impact of the Zeeman field is especially important for the
x=0.125 phase since at this particular composition the magnitude
of the in-plane superconducting gap is anomalously
low. \cite{dumm02}   The field-dependence of the $c$-axis
superfluid density within this scenario has the following form:
$\omega_s^2(H)/\omega_s^2(0)=\sqrt{1-(H/\Delta)^2}$ \cite{won} and
is displayed in Fig.3C for different magnitudes of the energy gap
$2\Delta$. Interestingly, the functional form of
$\omega_s(H)/\omega_s(H=0)$ is reproduced by this calculation. The
Zeeman reduction of the condensate strength occurs in concert with
other dissipation mechanisms and accounts for at least 10-15$\%$
depression of the $\omega_s(17 T)$ value. An additional factor
pertinent to the anomalous field response of the underdoped
crystals may be connected to spatial non-uniformities of
superconductivity {\it within} the CuO$_2$ planes revealed by a
variety of experimental methods. \cite{iguchi01}. It seems
plausible that the magnetic field will influence  coupling between
these dissimilar regions in the CuO$_2$ plane thus leading to
field-dependence of the {\it in-plane} superconducting parameters
such as $\lambda_{ab}$ in Eq.~\ref{eq:bulaevskii}.

Models of the superfluid density suppression discussed above
involve only a reduction of carrier density n$_s$. However, the
$c$-axis response of cuprates also reveals changes of the
effective mass m$^*$ or of the kinetic energy at
$T<T_c$.\cite{basov99} Recently Ioffe and Millis proposed that the
variation the effective mass below $T_c$ is connected with phase
coherence between the CuO$_2$ planes.\cite{ioffe-science} Phase
coherence is suppressed in magnetic field which may lead to a more
rapid degradation of the superfluid density via the increase of
$m^*$ in addition to usual reduction of $n_s$. This latter effect
is expected to be particularly strong in underdoped samples which
show the strongest changes of $m^*$ at $T<T_c$.\cite{basov99}

In conclusion, magneto-optical results for underdoped LSCO
revealed remarkable depression of the superfluid density in the
vortex state. We have identified several factors which may account
for much more complex behavior of underdoped LSCO beyond
conventional Josephson vortex models. Further theoretical analysis
is needed to distinguish between the roles played by spatial
non-uniformities of superconducting state, as well as by changes
of kinetic energy in the observed behavior. In this fashion
quantitative understanding of vortex state data presented in this
work will be instrumental in narrowing down the range of plausible
theoretical descriptions of the underdoped state cuprates.

We thank D.P.~Arovas, L.~Bulaevskii, Ch.~Helm, J.R.~Clem,
M.~Coffey, A.E.~Koshelev, T.~Koyama and A.J.~Millis for useful
discussions. The research supported by NSF, DoE, the Research
Corporation and the State of Florida.

\begin{references}
\bibitem{eta_values} These values are much smaller than typical
values measured previously for the vortex motion $across$ the
planes \cite{golosovsky94}, in qualitative agreement with a
theoretical prediction \cite{hao91}
$\eta_{\parallel}/\eta_{\perp}\approx \lambda_c^2/3\lambda_{ab}^2
\sim 1,000 - 3,000$. In the configuration studied here (see the
bottom inset of Fig.~\ref{fig:loss}) because of "intrinsic
pinning" \cite{doyle93} one intuitively expects viscous effects
(i.e. dissipation) to be much smaller than for the vortex motion
across the planes.
\bibitem{dumm02} Dumm M., {\it et al.}, {\it Phys.Rev.Lett.}
{\bf 88} (2002) 147003. IR measurements of the in-plane response
of the x=0.125 LSCO crystal show that the superfluid density is
collected from the frequency range below 70-80 cm$^{-1}$. This
sets up the upper bound for the magnitude of $2\Delta(0)$ whereas
the analysis of the $\omega$-dependence suggests that gap value
can be as small as 40 $cm^{-1}$.
\end{references}
\end{document}